\documentclass[onecolumn]{aastex631}

\usepackage{amsmath}
\usepackage{pifont}
\usepackage{selectp}
\usepackage{mathtools,amssymb}
\usepackage{multirow}
\usepackage[utf8]{inputenc}
\usepackage{nicefrac}
\usepackage{xfrac}
\usepackage{graphicx}
\usepackage{comment}

\usepackage[T1]{fontenc}

\newcommand\spinny{\texttt{SPINNY}}
\newcommand\multimoon{\texttt{MultiMoon}}
\newcommand\jt{$J_2$}
\newcommand\ct{$C_{22}$}

\def\kms{\ifmmode{\rm km\thinspace s^{-1}}\else km\thinspace s$^{-1}$\fi}

\def\x{\times}

\newcommand{\trackchange}[1]{#1}

\begin{document}

\pagenumbering{arabic}

\shorttitle{Non-Keplerian Binaries}
\shortauthors{Ragozzine, et al.}

\title{Beyond Point Masses. I. New Non-Keplerian Modeling Tools Applied to Trans-Neptunian Triple (47171) Lempo}

\author[0000-0003-1080-9770]{Darin Ragozzine}
\affiliation{Brigham Young University, Department of Physics and Astronomy, N283 ESC, Provo, UT 84602, USA}

\author{Seth Pincock}
\affiliation{Brigham Young University, Department of Physics and Astronomy, N283 ESC, Provo, UT 84602, USA}

\author[0000-0002-1788-870X]{Benjamin C. N. Proudfoot}
\affiliation{Brigham Young University, Department of Physics and Astronomy, N283 ESC, Provo, UT 84602, USA}
\affiliation{Florida Space Institute, University of Central Florida, 12354 Research Parkway, Orlando, FL 32826, USA}

\author[0000-0003-4051-2003]{Dallin Spencer}
\affiliation{Brigham Young University, Department of Physics and Astronomy, N283 ESC, Provo, UT 84602, USA}

\author[0000-0003-0333-6055]{Simon Porter}
\affiliation{Southwest Research Institute, 1050 Walnut Street, Suite 300, Boulder, CO 80302, USA}

\author[0000-0002-8296-6540]{Will Grundy}
\affiliation{Lowell Observatory, Flagstaff, AZ 86001, USA}

\email{darin\_ragozzine@byu.edu}

\setcounter{footnote}{0}

\begin{abstract} 
Many details of the formation and evolution of the solar system are best inferred by understanding the orbital and physical properties of small bodies in the solar system. For example, small body binaries are particularly valuable for measuring masses. By extending the models of small body binaries beyond point masses, new information about the shape and spin orientation becomes available. This is particularly informative for Trans-Neptunian multiples (two or more components) where shapes and spin orientations are poorly understood. Going beyond point masses requires modeling tools that no longer assume fixed Keplerian orbits. To this end, we have developed a new $n$-quadrupole integrator \texttt{SPINNY} (SPIN+N-bodY) and pair it with a Bayesian parameter inference tool \texttt{MultiMoon}, both of which are publicly available. We describe these tools and how they can be used to learn more about solar system small body multiple systems. We then apply them to the unique Trans-Neptunian hierarchical triple system (47171) Lempo, finding a three-point-mass solution for the first time. This solution has two surprises: unequal densities of the inner components and a dynamical configuration apparently unstable on the age of the solar system. 
\end{abstract}


\section{Introduction}
Solar system small bodies (SSSBs) provide unique and powerful insights into the formation and evolution of planetary systems. Their heliocentric orbital distribution is used to track the orbital configuration of the solar system planets over time. Their physical properties like size distributions, bulk densities, and spin orientations place limits on events in the early solar system like accretion properties. For those small bodies which are composed of multiple components -- binaries and higher-order multiple systems -- the orbital properties of these components are crucial for understanding how these systems were assembled. The formation stories for small body multiples then provide direct insights into the conditions of the solar system over time. And of course, the physical and orbital properties of solar system multiples are central to understanding the small bodies themselves including geophysical properties like tidal heating, surface insolation patterns, collision histories, etc. 

In the simplest case, an SSSB multiple is a binary which can be approximated as two point masses. Moving beyond point masses adds complications in identifying, characterizing, and understanding SSSB multiples; these complications are a blessing and a challenge. The complexity of dynamical interactions between multiple spinning bodies can illuminate properties that are otherwise nearly impossible to determine, but also leads to a more challenging process in precisely explaining observational constraints with specific models. The Trans-Neptunian Multiple (47171) Lempo (1999~TC$_{36}$) is an example of a complex system that has not been analyzed to its fullest extent because it requires a more complicated dynamical analysis. Though Lempo is a relatively unique in the solar system as a true hierarchical triple of three objects of similar masses, its orbital and physical properties have not been modeled self-consistently \citep{benecchi201047171}. Even without precise knowledge of the system, \cite{correia2018chaotic} was able to show that this system has rich spin-orbit-tidal dynamics. Furthermore, the frequency and properties of such triple systems are an important constraint on planetesimal formation \citep{nesvorny2010formation}. 

The non-spherical dwarf planet Haumea and its two moons and the four small moons orbiting around the Pluto-Charon binary are other examples where more complex dynamical interactions are required to properly model the observational data \citep[e.g.,]{rb09, 2015arXiv150505933P, 2023PSJ.....4..120P}. In the asteroid belt, there are several asteroids with multiple moons orbiting non-spherical primaries such as (87) Sylvia \citep{2001IAUC.7588....1B,2005IAUC.8582....1M}, (130) Elecktra \citep{2022A&A...658L...4B}, (216) Kleopatra \citep{2008IAUC.8980....1M}, etc. But even binaries with only two components may have informative spin-orbit dynamical interactions due to non-spherical shapes. 

Complex dynamical interactions take time to manifest observationally. In the past, dynamical models that used the point-mass assumption for binaries and/or that neglected gravitational interactions between small components were able to describe the vast majority of systems. However, the neglected non-Keplerian effects grow in time (usually $\sim$quadratically with the observational baseline) so that accurate explanation of future observations will require more complex models for many of the hundreds of known SSSB multiples going forward. 

With the goal of extracting unique and meaningful physical properties from SSSB multiples, we have developed tools to support more complicated non-Keplerian modeling of SSSB multiples. We begin by identifying the key physical properties neeeded to describe observations, particularly relative astrometry from direct imaging ($\S$2). Upon concluding that a new $n$-quadrupole integrator is essential for this modeling effort, we describe and validate our new \spinny{} integrator in $\S$3. We also discuss how the simplified computational model translates to real-world understanding of SSSBs. In $\S$4, we describe and validate \multimoon{}, our new Bayesian parameter inference engine for non-Keplerian modeling of relative astrometry. Both \spinny{} and \multimoon{} are publicly available at (\url{https://github.com/dragozzine/multimoon/releases/tag/v1.2}). We use these tools to  provide the first self-consistent non-Keplerian model of (47171) Lempo in $\S$5-7. Overall discussion and conclusions are then provided in $\S$\ref{sec:conclusions}. 

\section{Background}

Our goal is to provide new modeling tools that can account for detectable non-Keplerian effects. We thus begin with an overview of observational constraints ($\S$\ref{sec:observations}) and then investigate what physical processes are required to model them ($\S$\ref{sec:needquadrupoles}). 

\subsection{Observations of SSSB Multiples}
\label{sec:observations}
There are many ways to identify and characterize SSSBs with multiple components. The \href{https://www.johnstonsarchive.net/astro/asteroidmoons.html#1}{Johnston Archive} tracks a comprehensive list of known SSSB multiples \citep{2019pdss.data....4J}.

Large numbers of small asteroid binaries are discovered using lightcurves that show mutual events  \citep[transits/eclipses/occultations between components, see, e.g.,][]{2015aste.book..355M}. While valuable for gathering statistically large samples of binaries, lightcurve-based binaries are hard to characterize in greater detail and are not good candidates for modeling shape and spin effects. 

The most successful precise characterization method has been direct imaging of the SSSBs to measure precise relative astrometric positions and uncertainties. This is common for both large asteroids and Trans-Neptunian Objects (TNOs).  

While some wide (astrometric separation $\gtrsim$1'') TNO binaries can be imaged with 4m-class telescopes, most observations have been performed by 10m-class telescopes such as Gemini, Keck, VLT, and Subaru with the highest precision from the Hubble Space Telescope (HST). Due to its combination of cameras and wavelengths, the James Webb Space Telescope (JWST) is expected to be similarly precise as HST at relative astrometry, but only at the shortest wavelengths. 

Other (non-spacecraft) observations that can constrain the physical and orbital properties of SSSB multiples include: thermal imaging at longer wavelengths to obtain sizes and albedoes \citep[e.g.,][]{lellouch2013tnos}, resolved or unresolved photometric lightcurves \citep[e.g.,][]{sheppard2008photometric,rabinowitz2014rotational,2021Icar..35614098S}, resolved imaging of the primaries (especially for large asteroids and for the largest TNOs) \citep[e.g.,][]{stern1997hst,vernazza2021vlt}, stellar occultations \citep[e.g.,][]{ortiz2017size}, mutual events  \citep[e.g.,][]{rabinowitz2014rotational,2015aste.book..355M}, measuring absolute astrometry to constrain photocenter-barycenter motion (e.g., Tanga et al., in prep.) and occasionally other exotic phenomena. 

Ideally, all the observations about a particular object would be used in a single model to infer physical and orbital properties. Although Bayesian statistics provides the framework for combining different kinds of observational data (e.g., regularization), connecting the internal representation of the model to every kind of data is challenging. At present, we focus exclusively on relative astrometry, though we note that some data can be approximately represented that way, e.g., mutual events.

\subsection{Detectable Non-Keplerian Dynamics in SSSB Multiples}
\label{sec:needquadrupoles}

In real systems, there are many known non-Keplerian effects that can theoretically affect the time evolution of positions and spin orientations of orbiting objects: the ``permanent'' mass distribution of the components (e.g., permanent asphericity); the dynamic mass distribution (e.g., ``fluid'' rotational and tidal bulges); known and unknown additional perturbers, including those bound to the system as well as the Sun, planets, and other objects; tidal evolution; General Relativistic effects; non-gravitational forces and torques (e.g., from mass loss in comets); Yarkovsky/YORP effects; and even potentially electromagnetic forces. While it is useful to start with this more exhaustive list, in practice, we can neglect effects in our model that cause a systematic uncertainty below some desired threshold relative to the statistical uncertainty (e.g., <~10\% of the statistical uncertainty, $\sim$0.1 milliarcsecond relative astrometry) during the decades-long observational baseline. We also want to focus on detectable non-Keplerian effects which provide unique new information about SSSBs; this leads us to focus on Trans-Neptunian Multiples because they have orbits where non-Keplerian effects are observationally stronger and they generally have unknown shapes and spin orientations. \trackchange{TNO Multiples are also very powerful for detecting non-Keplerian effects because they have significant eccentricities (and potentially inclinations); non-Keplerian effects are very difficult to discern in the almost circular coplanar orbits of satellites around most well-studied asteroids.}

Considering possible causes of non-Keplerian effects and focusing on TNOs in the foreseeable future, we have concluded that the effects worth consideration are due to non-spherical shapes and multiple interacting components (whether bound or unbound, known or unknown). All of these configurations can be well approximated by a model that contains an arbitrary number of masses with fixed gravitational harmonics extended to quadrupole order ($J_2$ and $C_{22}$, defined below) whose three-dimensional orientations are tracked and coupled to the orbital motion through torques. Using the model in other circumstances necessitates confirmation that no additional dynamical effects should be modeled. 

These non-Keplerian effects are undetectable on timescales comparable to the orbital period(s) of the system, during which time relative motion is well-approximated as a fixed Keplerian orbit. The main observable effect of non-Keplerian dynamics, whether from non-spherical shapes or multiple interacting bodies, is a precession of the orbital orientation angles. The orbit normal changes direction (``nodal precession'') and the direction of periapse changes (``apsidal precession''). However, the amplitude of these effects may be small in practice. Nodal precession is only detectable if the torques are non-planar (e.g., if the spin pole of the objects are misaligned with the orbit pole). Apsidal precession is only detectable if the orbit has significant eccentricity since changes in the orientation of practically circular orbits is invisible. However, non-Keplerian motion is not fully described as simple orbital precession as there are additional short-term effects that we have confirmed are detectable in some cases. Robust detections result from measuring apsidal and/or nodal precession which has a period usually $10^{3-5}$ times longer than the orbital period. However, the precession rate can be well measured by SSSB Multiples which typical constrain orbital orientation angles with a precision of $\lesssim$3$^{\circ}$. This means that non-Keplerian motion can become readily detectable on timescales similar to current observational baselines of several years. For many known SSSBs, non-Keplerian motion will become an essential component of modeling orbital motion as new observations are gathered. 

We here briefly justify our choice to ignore effects beyond multiple quadrupoles with fixed quadrupole moments. For example, dynamical tidal bulges can be described by a time-varying quadrupole moment \citep[e.g.,][]{2009ApJ...698.1778R}, but for small bodies (and even for the Moon and Mercury) this component is negligible for spin-orbit dynamics (ignoring undetectably small librations) compared to the permanent asphericity. Rotational bulges (and tidal bulges in spin-orbit resonance) can be important in the sense that they cause the object to change shape especially over billion year timescales, but on observational timescales, these changes are considered part of the ``permanent'' asphericity that is already accounted for. Thanks to the nature of spherical harmonic expansion, higher-order harmonics (such as $J_4$) are effectively always smaller than the quadrupole and are not expected to be important for TNOs in the medium-term future. 

The gravitational harmonics associated with the stronger ``dipole'' term can be exactly canceled by choosing the center of each object to be at its center of mass. SSSBs have complex shapes and potentially heterogeneous internal structures, so that the center-of-mass may be displaced from the center-of-figure. Since our analysis focuses on interpreting relative positions in imaging data, this displacement is convolved with the issue of center-of-figure--center-of-light offsets. As a result, we need only consider the center-of-light--center-of-mass offset, which we call the photocenter-barycenter offset. (Note that we are discussing the offset for each resolved component, not the overall photocenter-barycenter variation of the full multiple system. However, even high precision imaging can lead to unresolved components that could have rather large photocenter-barycenter offsets.) That is, when this offset is assumed to be zero, the dipole term in the gravitational potential expansion vanishes. The photocenter-barycenter offset is an effect that is approximately included in our \multimoon{} model, but it does not need to be accounted for in the spin-orbit dynamics considered by \spinny{}. However, it is important to note that gravitational harmonics are calculated related to the center-of-mass, so that a center-of-mass--center-of-figure offset can also contribute to higher-order gravitational harmonics in addition to the usual effects from the shape and overall density distribution. 

We thus conclude that an arbitrary number of interacting quadrupoles -- which we call an $n$-quadrupole model -- is presently the ideal balance between complexity and accuracy; similar conclusions were reached in other work \citep[e.g.][]{2009CeMDA.104..103S}. \trackchange{This is likely to be true for the vast majority of SSSB binaries (not just TNOs) for the foreseeable future. The next improvement would probably be the inclusion of higher-order gravitational harmonics in well-studied asteroid binaries.}

\section{The \texttt{SPINNY} n-quadrupole integrator}

As discussed above, the next phase of SSSB multiple characterization requires tracking the positions, velocities, and spin orientations of multiple interacting components at the quadrupole level. One option is to focus on the orbital motion alone since the dominant observational strategy of relative astrometry cannot measure spin properties. However, for self-consistency and potential future modeling of other data (e.g., lightcurves), it is also valuable to include all the effects on both spin and orbital motion. In particular, this includes the ``back-torques'' of the orbital motion on the spins of the components which has not been included in other non-Keplerian SSSB fits in the past. 

Unlike long-term integrators, SSSB modeling also requires an emphasis on precise position information on decadal timescales representing $\sim$10$^{1-4}$ orbital periods. \trackchange{This implies that many dynamical packages interested in approximate long-term motion (such as symplectic integrators) are not ideal for our purposes.}

\trackchange{At the time of the inception of this project, there was no publicly-available efficient numerical n-quadrupole integrator. Therefore, we developed such an integrator to support the inference of non-Keplerian effects in SSSB Multiples which we call \texttt{SPINNY} (SPIN + N-bodY).}

\subsection{Justification for Numerical Modeling}
As discussed further below, the dominant non-Keplerian orbital effect is simply fixed precession rates. At first blush, adding these in as free parameters allows for a model where the osculating (e.g., instantaneous) orbital elements are easily determined at all times of observations and which can then be readily converted into relative positions. This suggests that a semi-analytical non-Keplerian model would be superior to a numerical integration of the n-quadrupole problem. This fails for three reasons. First, our detailed investigation shows that this simple parametrization is not precise enough to avoid systematic errors; that is, the observational data are of sufficient quality to go beyond this simple parametrization of the orbit as purely a uniformly-precessing ellipse. Second, more complex systems (such as two misaligned triaxial quadrupoles or two moons orbiting a triaxial primary) do not undergo such simple motion and instead have a complex interchange of angular momenta between multiple reservoirs \citep[][Proudfoot, in prep]{correia2018chaotic}. Finally, the conversion of orbital elements to spatial coordinates in the correct reference frame is relatively computationally expensive, making numerical integration competitive even for simpler problems. 

We thus turn to the numerical technique of ``integrating'' the differential equations describing an arbitrary number of gravitationally-interacting quadrupoles. Though an n-quadrupole integrator is only a few times more complicated than a standard n-body (e.g., n-point-mass) integrator, at the inception of this project, there was no publicly-available integrator designed for this problem. Though similar integrators are now available -- GRIT \citep[][]{chen2021grit}, SMERCURY-T \citep[][]{kreyche2021exploring}, and REBOUNDx \citep[][]{2023arXiv230300006L} -- they are not focused on fitting observational data for SSSBs and/or do not contain all the important physical effects. Developing our own integrator also allowed us to maximize its efficiency for the problem at hand: determining precise positions on $\sim$100 orbit timescales. For example, our short integration time-span means that we do not need to construct a symplectic integrator that would conserve energy on long integration timescales. 

\subsection{The \spinny{} Integrator}

We here present the new publicly-available \spinny{} integrator. \spinny{} (SPIN + N-bodY) is written in C++ to optimize speed, but additionally has a Python wrapper (using Cython) to enable easier usage. It uses a Cash-Karp 4(5)-order integrator with an adaptive timestep controlled by a user-specified tolerance \citep{cash1990variable}. \texttt{SPINNY} only calculates torques (of non-point-masses on objects and the equal and opposite back-torque on the spin orientation) when needed and defaults to n-body integration between point-masses when possible. Integration to specific times allows for quick querying of state vectors at the times of observations for efficient use within \texttt{MultiMoon}.

In addition to the usual integration of positions and velocities of each of the components, \spinny{} uses quaternions to track the orientations of each of the non-point-mass components, avoiding instabilities with integrating the Euler equations directly. The equations of motion are essentially identical to those presented in \cite{correia2018chaotic}. 

Internally, \texttt{SPINNY} tracks positions (in km), velocities (in km/s), quaternions (for three-dimensional spin orientation), and spin rate (in 1/s) using masses (in kg) and three body-fixed principal moments of inertia ($\mathcal{A,B,C}$ in kg km$^2$) for each object. While we could, in principle, use observational constraints to infer these parameters directly, we prefer to use more physically meaningful properties. 


\subsection{Other Quadrupole Representations}

\subsubsection{Gravitational Harmonics}

At the quadrupole level of approximation, principal moments of inertia are equivalent to the usual expansion of the gravitational potential ($U$) from an object with mass ($M$) at distance ($r$) in spherical harmonics \citep[e.g.,][]{yoder1995astrometric}:
\begin{equation}
    U(r,\theta,\phi) = -\frac{GM}{r}\left[1 - 
    J_2\left(\frac{R}{r}\right)^2\left(\frac{3}{2}\sin^2\theta - \frac{1}{2}\right) +
    C_{22}\left(\frac{R}{r}\right)^2\cos^2\theta\sin2\phi + 
    \mathcal{O}\left(r^{-3}\right)
    \right]
\label{eq:gravpot}
\end{equation}
where $J_2$ is the second-order zonal gravitational harmonic coefficient, $C_{22}$ is the second-order sectoral gravitational harmonic coefficient, $\theta$ is the body-fixed latitude-like angle, $\phi$ is the body-fixed longitude-like angle, and $R$ is typically defined as the volumetric radius. Note that this definition already includes placing the origin of the potential at the center of mass (thus eliminating dipole terms) which is discussed further below. It also assumes that $\phi$ is defined in such a way as to eliminate the $S_{22}$ term (so that the coordinate related to the rotation of the potential is zero with respect to the direction of the prolateness of the potential). 

The gravitational harmonics are related to the moments of inertia by 

\begin{equation}
\mathcal{A} = \frac{M}{5}(5 J_2R^2 - 10 C_{22}R^2 +2 c^2)
\end{equation}
\begin{equation}
\mathcal{B} = \frac{M}{5}(5 J_2R^2 + 10 C_{22}R^2 + 2 c^2)
\end{equation}
\begin{equation}
\mathcal{C} = \frac{M}{5}(10 J_2R^2+2 c^2)
\end{equation}

where $c$ is an assumed size, in km. In the triaxial ellipsoid approximation, it is the shortest semi-axis along the maximum moment of inertia. When actually calculating the gravitational potential and forces acting on bodies in the $n$-quadrupole problem, the quantities $J_2$ and $C_{22}$ are always combined with $R^2$ (similar to how mass is always coupled with $G$). We thus find it more meaningful and appropriate to solve for $J_2R^2$ and $C_{22}R^2$. The interpretation of $R$ also depends on the chosen shape model in a way that is independent from the non-Keplerian analysis and discussed further below.

\subsubsection{Shape Parameters for a Homogeneous Triaxial Ellipsoid}

The spin-orbit dynamics at the quadrupole level depends on the mass ($M$), oblateness ($J_2R^2$), prolateness ($C_{22}R^2$), and an effective size ($c$) for each component. (Practically speaking, it is difficult to measure all of these parameters even for one object.)

These parameters provide important, but non-unique, information about the shape of the objects. To infer specific shapes requires defining a shape model. Arguably the simplest and most common shape model for near-spherical objects is a homogeneous triaxial ellipsoid; we leave the discussion of other shape models for future work. At the quadrupolar level of approximation \citep[but not at higher order, see][]{2017CeMDA.127..369H}, specifying the moments of inertia or the gravitational harmonics are equivalent to determining the semi-axes of this ellipsoid, $a \ge b \ge c$. 

For a homogeneous triaxial ellipsoid with semi-axes $a \ge b \ge c$ approximated at the quadrupole order, these parameters are 
\begin{equation}
J_2R^2 = \frac{1}{5}\left(\frac{1}{2}a^2 + \frac{1}{2}b^2 - c^2\right)
\end{equation}
and 
\begin{equation}
C_{22}R^2 = \frac{1}{20}(a^2 - b^2)    
\end{equation}
following other derivations \citep[e.g.,][]{1994Icar..110..225S,yoder1995astrometric}.
\cite{correia2018chaotic} uses $J_2$ and $C_{22}$ alone and derives $C_{22} < \frac{J_2}{2}$ and $J_2 \le 0.2$. While $J_2$ defined this way may have an upper limit (since it assumes $R = \sqrt{2a^2 + 2b^2}$), $J_2R^2$ and $C_{22}R^2$ do not, which is more sensible in the case of extreme shapes or the effective gravitational potential of a contact binary or unknown component. In the case of an ellipsoid with $a \ge b \ge c$, we do still have $J_2R^2 \ge \frac{1}{2} C_{22}R^2$ with equality happening only when $b=c$, e.g., a body with prolateness but zero oblateness. This is typically enforced in \texttt{SPINNY}.
For example, \texttt{SPINNY} when given $c$, $J_2R^2$, and $C_{22}R^2$ determines $a$ and $b$ as: 
\begin{equation}
a = \sqrt{5J_2R^2 + c^2 + 10C_{22}R^2}
\end{equation}
\begin{equation}
b = \sqrt{5J_2R^2 + c^2 - 10C_{22}R^2}
\end{equation}
which is not meaningful when $J_2R^2 \ge \frac{1}{2} C_{22}R^2$; however, since \texttt{SPINNY} actually uses only moments of inertia, this constraint could, in theory, be relaxed.

\subsection{Application to Real Bodies}

Non-Keplerian modeling opens many new opportunities for improving our understanding of the physical and orbital properties of SSSBs. Measuring individual masses in binaries can lead to better density and albedo measurements. Shapes and spin orientations are particularly valuable for TNOs since they are otherwise difficult to measure for individual objects. Detection of unresolved components opens a large range of previously unavailable phase space for discovery. 

Still, there are several considerations to note when using the quadrupole approximation to understand SSSBs. \trackchange{To provide a thorough understanding of these considerations, we describe in this section observational limitations and degeneracies, shape modeling, additional components (bound and unbound, known and unknown), and the inference of spin poles, in particular.} 

\subsubsection{Observational Limitations and Degeneracies}

First, from an observational standpoint, it is difficult to get precise constraints on all the \spinny{} parameters. While in theory the integration depends on an effective size of each object ($c$), in practice, changing this has a very small effect on observational timescales since it only affects the strength of the back-torques and thus the angular momentum exchange on long timescales. Furthermore, in many SSSBs, the mutual orbit is much slower than the rotations of the individual components. In this case, \ct{} has little effect on the dynamics of the mutual orbit, as the contribution from \ct{} averages out. However, near spin-orbit resonances, which are seen in some systems, \ct{} can play a significant, even dominant role in the dynamics of the system (Proudfoot, in prep.). 

The dominant effect of non-Keplerian motion is to cause orbital precession. For a test particle orbiting around a body with only non-zero \jt, the apsidal and nodal secular rates can be written:
\begin{equation}
\label{eqn:apsidal}
    \dot{\omega} = -\frac{3n'J_2R^2}{2a'^2(1-e'^2)^2}\left( \frac{5}{2}\sin^2i'-2\right)
\end{equation}
\begin{equation}
\label{eqn:nodal}
    \dot{\Omega} = -\frac{3n'J_2R^2}{2a'^2(1-e'^2)^2}\cos{i'}
\end{equation}
\noindent where the test particle's orbit is described by the mean motion $n'$ ($\frac{2\pi}{P'}$ with $P'$ the orbital period), $a'$ is the semi-major axis, $e'$ is the eccentricity, and $i'$ is the inclination of the orbit relative to the quadrupole's ``equator'' ($\theta=0$ in Equation \ref{eq:gravpot}). Depending on whether apsidal and/or nodal precession are detected, different parameters can be inferred. Although non-Keplerian modeling can break degeneracies, such as measuring masses for both components in a binary, it also comes with new degeneracies, some quite strong. For example, if only nodal precession is detected, Equation \ref{eqn:nodal} shows that it difficult to disentangle the strength of $J_2$ from the angle between the spin pole and orbit normal ($i'$). Furthermore, in the simplest case, apsidal and nodal precession from different causes add linearly. This means that there are additional degeneracies when considering the effect of 2 quadrupoles (such as only being able to determine the sum of the $J_2R^2$ for both objects) or 1 quadrupole and precession induced from a third body \citep[as with Haumea, see][]{rb09}. In addition to using parsimonious parametrizations, any parameter inference model must be able to explore these degeneracies. 

\subsubsection{More Complex Shape Models}
Triaxial ellipsoids are often poor approximations to extended SSSB shapes such as contact binaries. In practice, measured values of mass, \jt{}, \ct{}, and other constraints must be interpreted within the context of a particular shape model \citep[][]{marchis2005mass} to provide specific size and shape estimates. Real objects may also be inhomogeneous, such as Haumea,  which is expected to be differentiated \citep[e.g., ][]{dunham2019haumea}. In this case, the relationship between the gravitational harmonics and the shape modeling is even more complicated (unlike for large ``fluid'' planets where \jt{} primarily provides information in the interior density distribution as a response to an external quadrupole field like rotation or tides, see, e.g., \citet{2009ApJ...698.1778R}). For SSSBs at the present level of precision, measuring heterogeneous density distributions is very challenging even in the best cases with a known shape. Without a known shape, the most parsimonious assumption is to attribute quadrupolar gravitational harmonics to the overall shape of the object unless detailed modeling can clearly establish heterogeneity. In other words, \jt{} and \ct{} cannot be readily used to measure differentiation for SSSBs like they are for larger objects. (As mentioned above, another contributor to gravitational harmonics beyond shape and density distribution is a center-of-mass--center-of-figure offset since gravitational harmonics are calculated relative to the center-of-mass.) 

In practice, the interpretation of \jt{} and \ct{} using homogeneous triaxial ellipsoids can still approximately represent information about the asphericity of SSSBs, so we defer detailed discussion of shape modeling to future work. Still, to emphasize the limitations of modeling SSSB shapes through non-Keplerian parameter inference, we use the more accurate term ``quadrupole'' (rather than ellipsoid) to refer to our model components. 

\subsubsection{Additional Unbound Components}
When considering the non-Keplerian effects due to additional components, it is helpful to distinguish between bound and unbound components (relative to the SSSB system being modeled). Unbound components potentially include the Sun, other planets, and potentially even passing SSSBs in extremely rare cases. The primary effect of these unbound components is a (somewhat time-varying) background ``tidal'' potential, with a strength characterized by the mass of the component compared to the mass of the SSSB system and the distance compared to the separation cubed. This implies that the Sun dominates all other effects as long as the small body is not within $\sim$1 AU of a giant planet (e.g., within the planet's Hill sphere), even for SSSBs in heliocentric resonances with planets (though potentially there are very long-timescale effects of resonant planets). While Kozai-Lidov-von Zeipel effects can significantly change orbits on long ($10^{3-5}$ yr) timescales \citep[e.g.,][]{naoz2010observed,porter2012kctf}, on observational timescales, the dominant effect from the Sun is precession of orbital orientation angles. The precession period is approximately the heliocentric orbital period squared divided by the SSSB mutual orbital period, i.e., the widest binaries (in terms of Hill spheres) are the most strongly effected. We have found that the effect is usually very small \citep[consistent with][]{2020MNRAS.494.2410E}, so our typical mode is to neglect the Sun in our calculations, but it can be included by adding it as a point-mass at the correct position and velocity within \spinny{}. 

\subsubsection{Additional Bound Known Components}
Additional known components (e.g., those included explicitly in the model) cause non-Keplerian effects even if all objects are point masses. 

Both non-spherical shapes and additional components primarily cause orbital precession. Indeed, to quadrupole order, the gravitational potential of two spherical components has an ``effective'' \jt{} and \ct{} of:
\begin{equation}\label{eqn:j2binary}
    J_{2}R^2 = \frac{1}{2} \frac{q}{\left(1+q\right)^2} a_s^2
\end{equation}
\begin{equation}\label{eqn:c22binary}
    C_{22}R^2 = \frac{1}{4} \frac{q}{\left(1+q\right)^2} a_s^2
\end{equation}
\noindent where $q$ is the mass ratio of the smaller to larger sphere and $a_s$ is the semi-major axis of the component's orbit. These equations account for the change in the center-of-mass with varying $q$. In this equation, we leave the combinations $J_{2}R^2$ and $C_{22}R^2$ since these are the physically meaningful parameters (see Equation \ref{eq:gravpot}).

Whether non-Keplerian motion is caused by another object or a quadrupolar potential, though the dominant effect of both is precession of the orientation angles, the difference can be detected with enough precise data comparable to the observational data currently available for some SSSB multiples. 

\subsubsection{Additional Bound Unknown Components}
Additional unknown (or, at least, unmodeled) components have the same dynamical effects discussed above. Since the primary effect of unknown components is orbital precession, a quadrupole model that uses \jt{} and possibly \ct{} can provide an initial approximation to the non-Keplerian perturbations whether it is due to shapes or unknown components. That is, modeling using \jt{} at first can reveal strong non-Keplerian motion which can then potentially be better explained by adding an unknown component. For example, an interior moon is a potential description of the observed motion of the orbit of Eris's moon Dysnomia (Spencer et al., in prep.). Detecting anomalously large values for \jt{} (e.g., $J_2 \gtrsim 0.5$ when assuming a triaxial shape model), may indicate that there are additional undetected system components. 

Non-Keplerian motion thus opens a powerful pathway to discovering unknown components of SSSB systems. This is similar to exoplanetary systems where planet-planet interactions (typically in the form of ``Transit Timing Variations'') have been used to infer and characterize the existence of planets solely based on their dynamical influence on observed planets. \citep[e.g., ][]{2012Sci...336.1133N}. Because multiple precise astrometric measurements require extensive imaging, these additional unknown components are expected to be either very faint or unresolved (or both). 

Detecting additional system components has the potential for gaining much deeper knowledge of the formation and evolution of a SSSB binaries. For large SSSBs with small moons, any additional components would likely be additional small interior moons. Uncovering the presence of these moons may indicate that the system was formed by collisional processes, like the small satellites of Pluto \citep[e.g., ][]{stern2006giant,bromley2020pluto}. If additional components are uncovered in small near-equal brightness binaries, it is possible that one of the system components itself may be an unresolved binary, forming a hierarchical triple system like the Lempo triple system. These systems are common outcomes of gravitational collapse simulations \citep[][]{nesvorny2010formation, robinson2020investigating, nesvorny2021binary}, although they may also be able to form through capture mechanisms \citep[][]{brunini2020origin}. 

We note that detection of an additional component from non-Keplerian motion is sensitive to a large range of parameter space that is poorly probed by other observational techniques. Lightcurves and occultations can sometimes detect whether a SSSB has an unknown companion, but these techniques can only distinguish a single object from two when the two components are similar in size and have very close (almost touching) orbits. This leaves a very large region where two components would be completely unresolved by any current telescopes, but where the two components are too far apart to reasonably detect in any other way. In this rather large phase space (approximately 3-30 primary radii), the only reasonably accessible observation is non-Keplerian orbital precession when the object is already in a binary or multiple system. Note that non-Keplerian orbital precession is actually stronger for wider separations (see Equation \ref{eqn:j2binary}), making it complementary to other methods that decrease in effectiveness with wider separations.  
 
\subsubsection{Determining Spin Poles} 
Most SSSB Multiples are large enough ($\gtrsim$few km) to expect fully principal axis rotation \citep[e.g.][]{2019MNRAS.485..725Q}. We implicitly assume in our definitions that the smallest semi-axis $c$ is aligned with the rotation pole of the body. Due either to higher-order terms or non-principal axis rotation (which \texttt{SPINNY} can integrate), it is possible in theory for the ``spin'' orientation angle to actually be just the orientation angle relevant to the effective quadrupole and to actually be slightly misaligned with the true spin pole. We neglect that small and currently undetectable effect. 

As is clear from Equations \ref{eqn:apsidal} and \ref{eqn:nodal}, the non-Keplerian motion induced by the gravitational harmonics require knowledge of both the strength of the gravitational harmonics \textit{and} the direction of the spin axis (to determine the appropriate $\theta$ and $\phi$ values in Equation \ref{eq:gravpot} or to determine the appropriate inclination in equations \ref{eqn:apsidal} and \ref{eqn:nodal}). Thus detection of non-Keplerian effects allows constraints to be placed on both SSSB shapes and spin poles.  

While many asteroids have known rotation poles, the rotation poles of TNOs have proved to be difficult to constrain due to their long heliocentric orbital periods. Even with the combination of lightcurves, occultations, and thermal measurements, measuring precise spin pole directions is not always possible. Non-Keplerian modeling provides a new tool for spin pole constraints with implications for studies of binary formation and evolution, insolation patterns, and the interpretation of other observations (such as lightcurves).

\subsection{Validation of \texttt{SPINNY}}
We have performed multiple tests to ensure that \texttt{SPINNY} is accurately integrating the n-quadrupole problem. Since \texttt{SPINNY} includes all the forces and torques relevant to the n-quadrupole problem, we begin by confirming that it conserves overall energy and angular momentum at a level well below observational precision. As \spinny{} is not a symplectic integrator, it does have slow drifts in energy and angular momentum that are related to the integrator tolerance. (In practice, we use \multimoon{} on a case-by-case basis to test likelihood calculations with many different tolerance values to determine the tolerance at which systematic errors from imprecise integration become negligible compared to the statistical uncertainties from the relative astrometry.)

\begin{figure}
    \includegraphics[width=\linewidth]{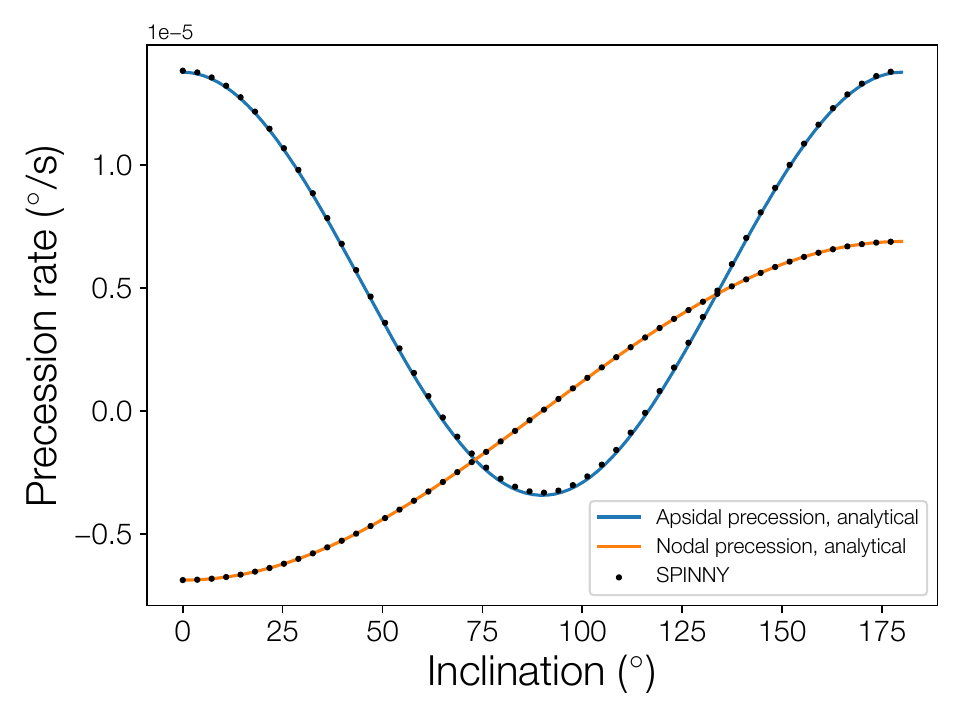}
    \caption{Comparison of analytical precession rates against \texttt{SPINNY}. Here, we simulate a series of massless test particles in orbit around the Earth ($a = 14400$km, $e = 0.1$, $\omega = 90\degr$, $\Omega = 90\degr$, $\mathcal{M} = 0\degr$, $J_2 = 0.001083$) with varying inclinations (w.r.t the Earth's equator) over 100 orbits. The observed precession rate in the simulated orbits (black circles) are compared to predicted second-order (in eccentricity) analytical precession rates of the system (solid blue and orange lines). The close agreement between the simulations and analytical predictions helps validate \texttt{SPINNY}. The effects of higher-order terms (some of which are present in \texttt{SPINNY}, but not the analytical equations) can be seen in the slight differences between the \texttt{SPINNY} simulations and the analytical predictions, especially for apsidal precession.}
    \label{fig:validation}
\end{figure}

As shown in Figure \ref{fig:validation}, \texttt{SPINNY} accurately matches analytical results for apsidal and nodal precession rates for an oblate primary. We also build a simulation of the chaotic spin of a small prolate secondary in a highly eccentric orbit described in \citet{2000ssd..book.....M} Section 5.4 and confirm that we retrieve the same orientation as a custom integrator of the equation therein. 

We have developed an analytical model of a massless small body in an arbitrary orbit around an oblate and prolate primary which has demonstrated remarkable agreement with \texttt{SPINNY}, supporting both models. 

Finally, \texttt{SPINNY} and \multimoon{} together give reasonable and plausible results that match previous results and our dynamical expectations. For example, we have reproduced the results for the three-body fit of Haumea and its two moons from \citet{rb09}. We used REBOUNDx \cite{2020MNRAS.491.2885T} to generate synthetic astrometry for an object orbiting a quadrupole and used  \multimoon{} powered by \spinny{} to correctly recover the orbital and physical parameters of the system. We have calculated the spin dynamics of Haumea's outer moon Hi'iaka and shown that it appropriately matches the analytical discussion in \citet{2016AJ....152..195H}.

Through the combination of these tests, we consider \texttt{SPINNY} to be validated and ready for general use. It is publicly available as part of \multimoon{}. 

\section{\multimoon{}: Bayesian Parameter Inference for non-Keplerian SSSB Astrometry}

Thanks to \texttt{SPINNY}, the relative positions between any pair of $n$ quadrupoles can be determined at any time given the required orbital, spin, and physical properties at a particular epoch. That is, \texttt{SPINNY} can calculate the forward model for the observed positions of components in an SSSB multiple given the orbit, shape, and spin parameters that we desire to infer. The relative positions between the ``primary'' object and other components are then projected into the plane of the sky at the time of observation using the geocentric ecliptic J2000 longitude and latitude at the time of the observation to determine precise relative astrometric positions. The ephemerides of the SSSB system (relative to the Earth) are taken from JPL Horizons, as queried by \texttt{astroquery} \citep[][]{ginsburg2019astroquery}. These ephemerides are corrected for light-time variations and other astrometric aberrations due to Earth's orbital velocity. We choose to account for light-travel effects by incorporating them directly into the observational data files; that is, \multimoon{} takes in data that is based on a clock co-moving with the SSSB. These data also sometimes requires a rotational transformation of both observations and uncertainties from equatorial coordinates to ecliptic coordinates. At present, \multimoon{} assumes the observations are taken from the Earth geocenter, which induces negligible systematic uncertainties for observations from Earth's surface or HST. However, for JWST, this could induce small systematic errors for certain viewing geometries. Fortunately, this could be easily accounted for by using JWST's ephemeris from JPL Horizons. Similarly, observations from other locations could be appropriately modeled by using the correct ephemeris.

To support efforts in the practical completion of the parameter inference problem, we have developed and publicly shared \texttt{MultiMoon}. Following recent trends in astrostatistics and with a desire to produce robust outputs, \multimoon{} uses Bayesian statistics to perform its parameter inference. Bayesian inference is naturally good at exploring parameters that have non-Gaussian posteriors (e.g., eccentricities near zero), degeneracies (which are more common when including non-Keplerian parameters), and/or weak constraints. 

\multimoon{} was desired to support the user in easily exploring complex dynamical fits to SSSB astrometry based on previous experience with this problem \citep[][]{rb09} and similar model-fitting problems \citep[e.g.,][]{macdonald2016dynamical,proudfoot2019modeling}. It is designed to enable moderately-trained undergraduates to perform meaningful scientific investigations. In the implementation of \multimoon{}, we have sought to follow the guidelines in the WAMBS (when to Worry and how to Avoid the Misuse of Bayesian Statistics) checklist \citep[see][]{van2021bayesian}. 

At its core, \multimoon{} uses Bayesian parameter inference to solve the orbit fitting problem. Bayesian inference requires the calculation of a likelihood function given a set of parameters and observed relative astrometry data, as well as prior distributions. Customized prior probability distribution functions for each parameter can be specified by the user, however, a flat, uninformative prior is most commonly used. The likelihood function currently used in \multimoon{} makes use of the chi-square statistic (e.g., each observation is independent with given fixed Gaussian uncertainties) as a default, which is commonly used in most types of model fitting problems. 


\multimoon{} uses the \texttt{emcee} ensemble Markov Chain Monte Carlo (MCMC) method to sample from the \emph{posterior} distribution \citep[][]{foreman2013emcee}. An ensemble of walkers explores the parameter space simultaneously, using the distribution of walkers to propose efficient moves that follow the likelihood surface. \multimoon{} begins with a user-specified ``burn in'' (or warm-up), after which a clustering algorithm \citep[based on][]{hou2012affine} removes highly underperforming walkers stuck in local maxima. The removed walkers are replaced with random linear combinations of existing walkers (which then undergo another short burn in). This pruning process is relatively unusual among similar implementations of \texttt{emcee} and is effectively a complex method for selecting the initial positions of walkers that significantly improves the performance of \multimoon{}, without jeopardizing the final results of the fits.

The ensemble is then propagated forward a user-specified number of steps to produce the final ``chain'' of all the parameters for each (post-burn-in) walker-step. At the end of the ``run,'' Various convergence diagnostics are computed in the form of plots and metrics and the user can start a new run readily from the end of a previous run. For a complex dynamical analysis, our ``runs'' of \multimoon{} typically use hundreds to thousands of walkers with a burn-in of $\sim$thousands of steps and a final sample of $\sim$thousands of steps.

Based on the output posteriors, \multimoon{} produces a variety of different output plots, including both MCMC diagnostic plots (e.g., walker trace plots, likelihood plots, etc.) and near-publication-ready plots \citep[e.g., corner plots,][]{foreman2016corner}. It also produces plots of astrometric fits and residuals to allow the user to inspect the fit and posteriors. 

Bayesian parameter inference is computationally expensive since it can easily require millions or even billions of likelihood evaluations. When \multimoon{} detects that the model is based on two point masses, it uses a standard Keplerian model. This is powered by the python \texttt{spiceypy} package \citep{annex2020spiceypy}, a Python implementation of SPICE \citep[][]{acton1996ancillary} and is very fast ($\sim$0.01 CPU-seconds per likelihood evaluation). When non-Keplerian effects are to be modeled, even our relatively optimized \spinny{} code takes about 0.1-1 CPU-seconds per likelihood evaluation. In combination with the increased number of parameters (and thus larger number of steps to achieve convergence of the MCMC to the posterior), non-Keplerian parameter inference can be very computationally expensive. 
Thus, \multimoon{} is parallelized to take advantage of the ensemble-nature of \emph{emcee}. For example, at each step, a walker can be assigned to a single core, allowing for a speed-up that is nearly linear in the number of cores (up to the number of walkers). Heuristically, though the total number of walker-steps to sample the posterior is similar, convergence is somewhat faster for many walkers taking fewer steps. We commonly use 960 cores on \multimoon{} (chosen to match our computational architecture) so that even complex analyses only take $\sim$1 day of wall clock time.


Like all parameter inference, \multimoon{} is more efficient when initial guesses for all the walkers are near the maximum likelihood. Initial guesses can be drawn from a multi-dimensional Gaussian or from a previous run. \multimoon{} also has a mode where a standard optimizer (instead of posterior sampler, see discussion in \cite{foreman2013emcee} \& \cite{2016PhT....69f..59H}) on all the initial guesses can be used. This is done with the popular \texttt{scipy.optimize.minimize} optimizer \citep[][]{2020SciPy-NMeth}. The user may select a specific method of optimization, but the default and recommended optimization method in this case is the Nelder-Mead simplex algorithm. This can be used to quickly explore parameter space, especially at the beginning of a new analysis. 

There are many equivalent ways to represent the parameters within \multimoon{} and the ensemble nature of \emph{emcee} is relatively robust to linear transformations. To help optimize the fitting process, we chose a parametrization that focused on observable quantities that can then be extended to desired physical quantities as ``derived'' parameters. (This is straightforward in Bayesian parameter inference because each sample from the chain is a self-consistent sample from the \emph{posterior}.) \multimoon{}'s parameters, their variable names, and units are provided in Table \ref{tab:mm_params}. Orbital and Physical parameters are provided at a specified epoch near the center of the data, however, in practice, the epoch used is usually chosen to match that used in previous studies. 

While in theory all SSSBs are described by quadrupoles, in practice not all of these parameters can be inferred from existing observational data. For example, the prolateness of rapidly rotating objects effectively averages out so that $C_{22}$ is nearly impossible to infer (Proudfoot, in prep.). As a result, each of the $n$ objects modeled by \multimoon{} can be considered as either a point mass, an oblate object, or a oblate and prolate object, which we indicated using ``Dynamics Flags'' of 0, 1, and 2, respectively. Oblate and prolate objects must have a specified spin pole (at epoch) and prolate objects must have a specified direction of orientation of the smallest moment of inertia axis at epoch (see Equation \ref{eq:gravpot}). For ease of use, instead of using a pole and zero-longitude direction vector in ecliptic coordinates, we use Euler angle rotations to represent the transformation of the coordinate system from the ecliptic frame to the body-fixed frames. We elect to use the same Euler angle rotations that describe orbital orientation angles and call these $\omega^{sp}$, $i^{sp}$, and $\Omega^{sp}$. These angles are directly related to the $\theta$ and $\phi$ coordinates of the quadrupole potential in Equation \ref{eq:gravpot}: $\theta$ is the latitude-like angle measured with respect to the potential axis determined by $i^{sp}$ and $\Omega^{sp}$ and $\phi$ is the longitude-like angle defined so that $\omega^{sp}$ is aligned with the minimum moment of inertia (so as to eliminate the ``$S_{22}$'' term in the potential). With the very reasonable assumption of principal axis rotation, $i^{sp}$ and $\Omega^{sp}$ describe the direction of the rotational pole (or orbit pole when approximating an unresolved component) and $\omega^{sp}$ describes the direction of the long-axis. Using the usual spherical trigonometry, it is possible to determine the mutual inclination ($i_{mut}$) between the potential axis and the orbital axis: $\cos i_{mut} = \cos i \cos i^{sp} + \sin i \sin i^{sp} \cos(\Omega - \Omega^{sp})$. This mutual inclination is a valuable constraint on the formation of the binary. 

\begin{table}[]
    \centering
    \textbf{Parameters for \multimoon{}}\par\medskip
    \begin{tabular}{l c c c c}
    \hline
        Parameter & Variable & Units & Name & Dynamics Flag(s) \\
        \hline
        mass & $M$ & kg & \texttt{mass\_i} & 0,1,2 \\
        semi-major axis & $a$ & km & \texttt{sma\_i} & 0,1,2\\
        eccentricity & $e$ & ... & \texttt{ecc\_i} & 0,1,2\\
        inclination & $i$ & deg & \texttt{inc\_i} & 0,1,2\\
        argument of periapse & $\omega$ & deg & \texttt{aop\_i} & 0,1,2\\
        longitude of ascending node & $\Omega$ & deg & \texttt{lan\_i} & 0,1,2\\
        mean anomaly & $\mathcal{M}$ & deg & \texttt{mea\_i} & 0,1,2\\
        oblateness & $J_{2}R^2$ & km$^2$ & \texttt{j2r2\_i} & 1,2 \\
        $i$-like spin orientation angle & $i_{sp}$ & deg & \texttt{spinc\_i} & 1,2\\
        $\Omega$-like spin orientation angle & $\Omega_{sp}$ & deg & \texttt{splan\_i} & 1,2\\
        spin-rate & $spin$ & rad/sec & \texttt{sprate\_i} & 1,2\\
        prolateness & $C_{22}R^2$ & km$^2$ & \texttt{c22r2\_i} & 2 \\
        $\omega$-like spin orientation angle & $\omega_{sp}$ & deg & \texttt{spaop\_i} & 2\\
     \end{tabular}
    \caption{Variables used by \multimoon{} and \spinny{} to produce orbital models. The dynamical flags used to define the shape of objects are included, and the corresponding variables of that model are shown. All angles relative to ecliptic J2000.}
    \label{tab:mm_params}
\end{table}

The parameters that are used in \multimoon{} are given in Table \ref{tab:mm_params}. These parameters are designed to be easy for the user to interface with; other derived parameters can also be constructed using the posterior distribution. However, internally \multimoon{} uses different variables and units which make no difference in the final results, but which speed up MCMC convergence. Due to the degeneracies in these units, we internally perform an exact coordinate transformation (and its reverse) and re-scaling to more efficiently sample parameter space. 

Based on experience with these problems \citep{RH10}, we use equinoctial elements which are closely related to the Thiele-Innes constants used for astrometric orbits of stellar binaries. These elements reduce degeneracies because they are measured relative to specific locations in space (longitudes) instead of relative to other orbital elements (arguments and anomalies). Semi-major axis $a$ is not affected, but the other orbital elements are transformed to $\lambda \equiv \mathcal{M} + \varpi \equiv \mathcal{M} + \omega + \Omega$, $e \sin \varpi$, $e \cos \varpi$, $\tan(i/2) \sin \Omega$, and $\tan(i/2) \cos \Omega$. These elements also remain well-defined as eccentricity and inclination (relative to the ecliptic) go to 0. 

Kepler's Third Law, when $J_2 = 0$, connects the well-observed Period and semi-major axis to the total of all mass interior to the orbit. This is often handled by using Jacobian orbital elements that are referenced to the center of mass of all interior objects. We elect not to use Jacobian orbital elements, even internally, but approximate the effect of Jacobian elements by transforming our coordinates to fit $m_1$, $m_1+m_j$, $m_1+m_j+m_k$, etc., where the bodies (after the primary) are chosen in order of semi-major axis. 

There is also an option to ``lock'' the spin angles for a specific body $j$, so that it sets $\omega_j = \omega_j^{spin}$, $\Omega_j = \Omega_j^{spin}$, and $i_j = i_j^{spin}$, at epoch. Spin locking for the primary is relative to object 2. This is an approximation in the case of multiple bodies, but has the effect of forcing obliquity to be $\sim$0 and the long-axis to point $\sim$at the primary at periapse, allowing for an approximate exploration of the effect of $J_2$ and $C_{22}$ without adding as many free parameters. Spin-orbit synchronization is more complex and has not yet been investigated.

\subsection{Validation of \multimoon{}}

We have applied \multimoon{} to many synthetic and real TNO systems in the process of testing, exploring, and validating. Keplerian fits to all known TNO binaries retrieve parameters that are statistically the same as published results (Proudfoot et al., in prep.). Fits to the Haumea system produce the same results as previous analyses based on codes used for \citet{rb09}. Non-Keplerian fits have already produced many interesting and plausible results that will be the discussion of future work. The reproduction of published work and overall consistency of \multimoon{} demonstrate that it is reasonably validated. 

\section{Application of \texttt{SPINNY} and \multimoon{} to Lempo}

The ``trinary'' or hierarchical triple configuration of the (47171) Lempo 1999 TC$_{36}$ system (an equal-mass inner binary orbited by an nearly-equal-mass moon) is unique in the solar system. Such configurations are expected from the streaming instability model for TNO binary formation \citep[][]{robinson2020investigating,nesvorny2021binary} and can help constrain that process. Furthermore, the complex dynamics of the Lempo triple system, such as those explored by \citet{correia2018chaotic} can provide insight into the density, spin orientations, and tidal properties of TNOs. 

Currently, the best solution for the orbital properties of the Lempo system is from \citet{benecchi201047171} which approximate the motion of the bodies as an inner Keplerian orbit and an outer Keplerian orbit around the center-of-light of the inner binary. After reviewing the physical, spin, and orbital characteristics of this system, we apply \multimoon{}, powered by \spinny{}, to this system to infer improved orbital and physical properties. 

\subsection{The Lempo Triple System: Physical Characteristics}



Lempo's heliocentric orbit is in the 3:2 ("plutino") mean motion resonance with Neptune: semi-major axis of 39.2, orbital period of 246 years, eccentricity of 0.22, and inclination of 8.4$^{\circ}$. The hierarchical nature of the system that primarily points to the streaming instability highlights the ongoing investigation into how the hot TNO population originally formed in the primordial disk. 

\citet{2021PSJ.....2...10F} combine visible and infrared measurements of Lempo to propose a composition of 30 ± 20\% H$_2$O, 20 ± 10\% silicates, and 50 ± 10\% complex organics, which is in approximate agreement with more detailed spectral models \citep{2003Icar..162..408D,2005A&A...444..977M,2008AJ....135...55B,2009Icar..201..272G,2015Icar..252..311D}. Various combinations of tholins, serpentine, water ice, carbon ices, and amorphous carbon have been proposed \citep[e.g.][]{2009A&A...501..375P,2015Icar..252..311D} and should be significantly clarified by JWST NIRSpec IFU observations taken in January 2023 (Program 2418, PI: Pinilla-Alonso). 

HST observations in 2001 showed a companion moon, now known as Paha, at about 0.36'' ($\sim$8700 km) away that was 16\% as bright as the ``primary'' object. Careful investigation with the best HST data from the Advanced Camera for Surveys High Resolution Camera (HRC) showed that the primary appeared slightly elongated which led to the discovery of the inner binary named Lempo-Hiisi. Point Spread Function fitting with HRC by \citet{benecchi201047171} were able to corrobate the existence of all three bodies, finding separations as small as 1.5 pixels ($\sim$0.03''), roughly one half of HST's diffraction limit. Lempo and Hiisi are very similar in brightness, but the orbital solution from \citet{benecchi201047171} does uniquely disambiguate each object at each epoch.

Table \ref{tab:fit_results} gives the orbital elements calculated by \citet{benecchi201047171} which are based on a double-Keplerian model. While these Keplerian parameters are a qualitatively good fit, the $\chi^2$ goodness-of-fit metric shows that the double-Keplerian model is statistically unacceptable. For example, the total mass of the system yielded two contradictory values: a total mass of $M_{sys}$ = 14.20$\pm${0.05}$\times10^{18}$ kg for the Lempo–Hiisi binary solution, but $M_{sys}$ = 12.75$\pm${0.06}$\times10^{18}$ kg for the outer binary. 
 
\begin{figure}[h!]
    \centering
    \includegraphics[scale=0.45]{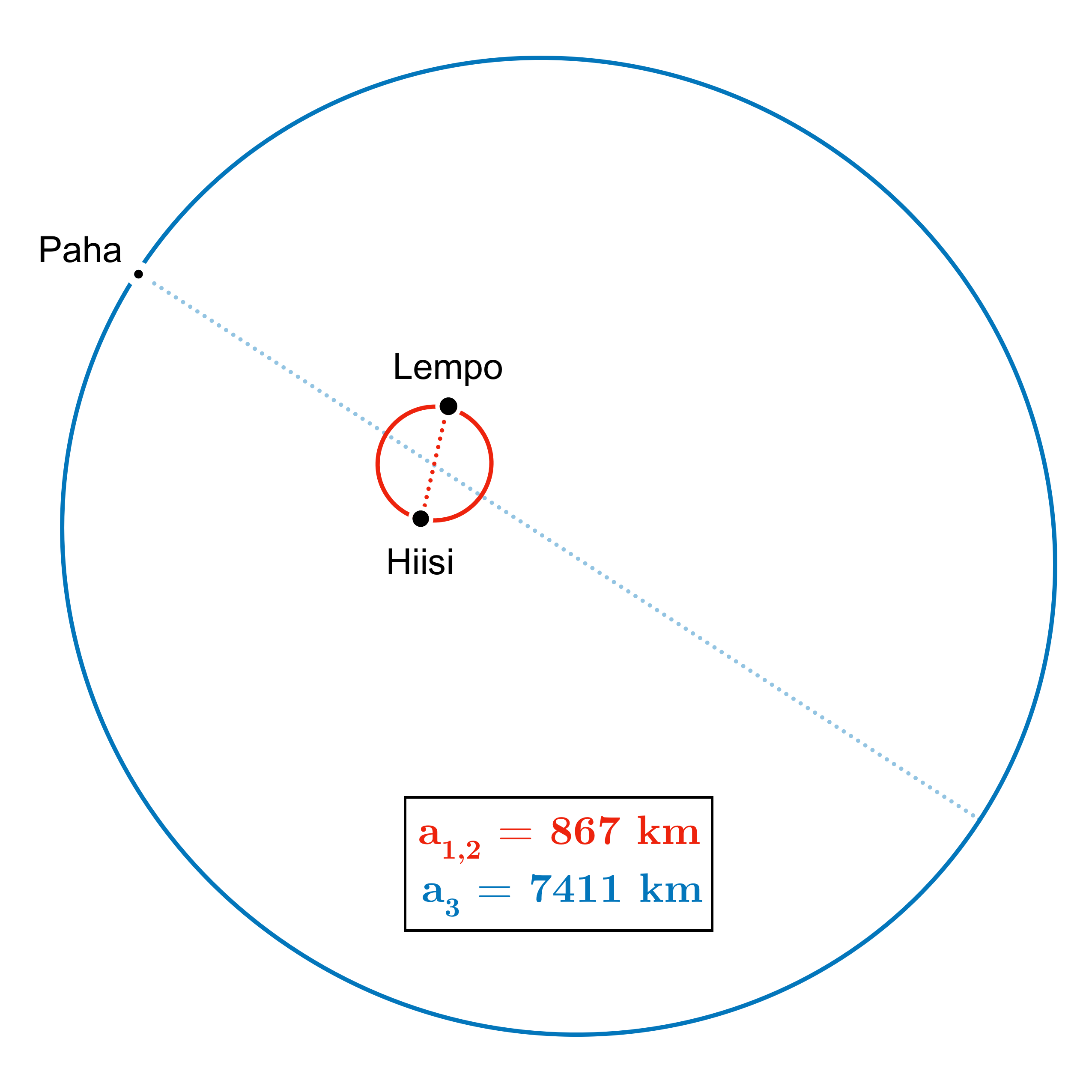}
    \caption{Configuration of the orbits of the three bodies in the Lempo triple system according to Benecchi, et al (2010). The joint orbit of Lempo and Hiisi is shown in the middle (red) and Paha's orbit is to the outside (blue). The dotted lines indicate the orbits' lines of apses. The black dots are included to show the relative sizes of the bodies and their separation. Note that a mutual inclination of $\sim$11$^{\circ}$ is not shown. }
    \label{fig:orbits}
\end{figure}

\citet{mommert2012tnos}, using thermal imaging from Herschel (in agreement with previous models, see Stansberry et al. 2008) and assuming equal albedoes of 0.079$^{+0.013}_{-0.011}$, derived an estimate for the diameters of the three bodies in the system: $272^{+17}_{-19}$ km for Lempo, $251^{+16}_{-17}$ km for Hiisi, and $132^{+8}_{-9}$ km for Paha. 

\subsection{Spin Characteristics}
\trackchange{Due to their close-proximity orbits, it is extremely likely that the Lempo–Hiisi binary is mutually tidally locked; they have a despinning timescale of $\sim$10$^{7}$ years even given weak tidal parameters \citep{2016AJ....152..195H}. Paha's perturbations would only serve to increase tidal damping and accelerate the progression towards the circular doubly-synchronous state.} 
On the other hand, the motion of the inner binary may prevent Paha from becoming tidally locked. Comparable circumstances have been observed with the moons of Haumea \citep{2016AJ....152..195H} and the small moons of Pluto \citep{2016Sci...351.0030W}. Free-spinning, rapidly-rotating bodies in the Lempo are plausible, especially for Paha. 

One study \citep[][]{ortiz2003study} suggested a rotational period of 6.21$\pm{0.2}$ hours for Lempo. This equates to roughly 7.4 rotations within Lempo–Hiisi's 1.9-day binary orbital period. However, this estimate was based on a low-amplitude, fragmentary lightcurve and was not able to be confirmed with previous data nor with any subsequent runs \citep[e.g.,][]{benecchi201047171}. The bodies of the Lempo–Hiisi binary are not well-resolved, and so their lightcurves are likely complex and difficult to decipher without a substantial number of new observations. However, they do show some variations indicating shapes that are non-spherical at the $\sim$5\% level.

\subsection{Orbital Characteristics}
Lempo's trinary configuration of three near-equal masses is unique in the solar system, though it is common in stellar triples where it has been well-studied. It is often approximated, as in the fit by \citet{benecchi201047171} as an inner Keplerian binary and an outer object which follows a Keplerian orbit around the center of mass of the inner binary. Such configurations of three equal-masses are only dynamically stable when the ratio of the periapse of the outer binary to the apoapse of the inner binary is $\gtrsim$10 as with Lempo ($\sim$5.8). Instability would lead to a reconfiguration into a system that is more dynamically stable either through ejection or accretion. Since Lempo's outer binary is so close to the inner binary, the non-Keplerian effects on both orbits are expected to be quite high.

All three bodies have small orbits when considering semi-major axis measured in radii; the Lempo—Hiisi binary which is only separated by approximately 5.5 Lempo radii. In addition, all three bodies are small enough (radii $\lesssim$ 150 km), that they likely have irregular shapes like similar size objects throughout the solar system which have estimated $J_2$ values near $\sim$0.1 \citep[e.g.][]{2016AJ....152..195H}. This implies that spin-orbit dynamics will be important for the Lempo system, though the shapes and spins of the components are not yet known. 

Using the approximate orbital solution from \citet{benecchi201047171}, \citet{correia2018chaotic} performed detailed investigation of the spin-orbit-tidal dynamics of the Lempo trinary for a variety of possible shape and spin properties. The dynamics were studied at the quadrupole-level of approximation using an integrator very similar to \spinny{} on 10$^{3-5}$ year timescales, additionally including a model for tidal evolution. The main result was that the Lempo system is highly chaotic and, with parameters from \citet{benecchi201047171} was dynamically unstable, dissolving in $<$1 MYr. This was mainly caused by the high eccentricity (0.1) of the inner binary leading to chaotic spin and orbit evolution for practically any shape and a wide range of spin periods. This is consistent with similar well-known results for the chaos of Hyperion \citep{wisdom1984chaotic} which also has an eccentricity of 0.1. Changes in torque on the irregular-shaped object during the orbit leads to strong, overlapping spin-orbit resonances which leads to chaotic spin. The close separation of these similar-sized objects produces torques that strongly couple spin and orbital properties, so that the spin chaos leads to orbital chaos and instability. For Lempo-Hiisi in particular, \cite{correia2018chaotic} found that the eccentricity of 0.1 caused the system to go completely unstable in $<1$Myr, even for shapes far more spherical than allowed by light curve data, and even when including tidal dissipation. A low eccentricity could also be expected because the tidal circularization timescale is 
$< 1000$\,yr; though this is based on a simple two-body circularization argument which is not directly applicable in a potentially chaotic trinary system, it is still $\sim10^7$ times shorter than the age of the Solar System. 

The existence of the Lempo system at the present time (and the lack of any plausible argument for why it would be in a configuration unstable on a timescale $\sim$10000 times shorter than its expected age) suggests that the analysis of \citet{correia2018chaotic} is not representative of the real Lempo system. He suggested that this was because the orbital elements were approximated by \citet{benecchi201047171} and that more detailed analyses of the astrometry with a more accurate model would presumably lead to a very different configuration that would be stable. With \multimoon{}, we can test this hypothesis directly. 


\subsection{\multimoon{} Fits}

Using \multimoon{}, we fit a three-point-mass model to the orbits of the Lempo-Hiisi-Paha system. The fit used 3840 walkers with a burn-in of 20000 steps, 5000 post-clustering burn-in steps, and sampling of 10000 steps. 
Convergence of the fit was assessed visually: walker plots showed adequate mixing of walkers and likelihood plots showed smooth surfaces, indicating that the posterior distribution has been adequately explored. We also confirmed that the \spinny{} tolerance was adequate at the level of precision relevant to the observational data. 

\section{Results}
\begin{figure}[p]
    \makebox[\linewidth]{
        \includegraphics[width=1.15\linewidth]{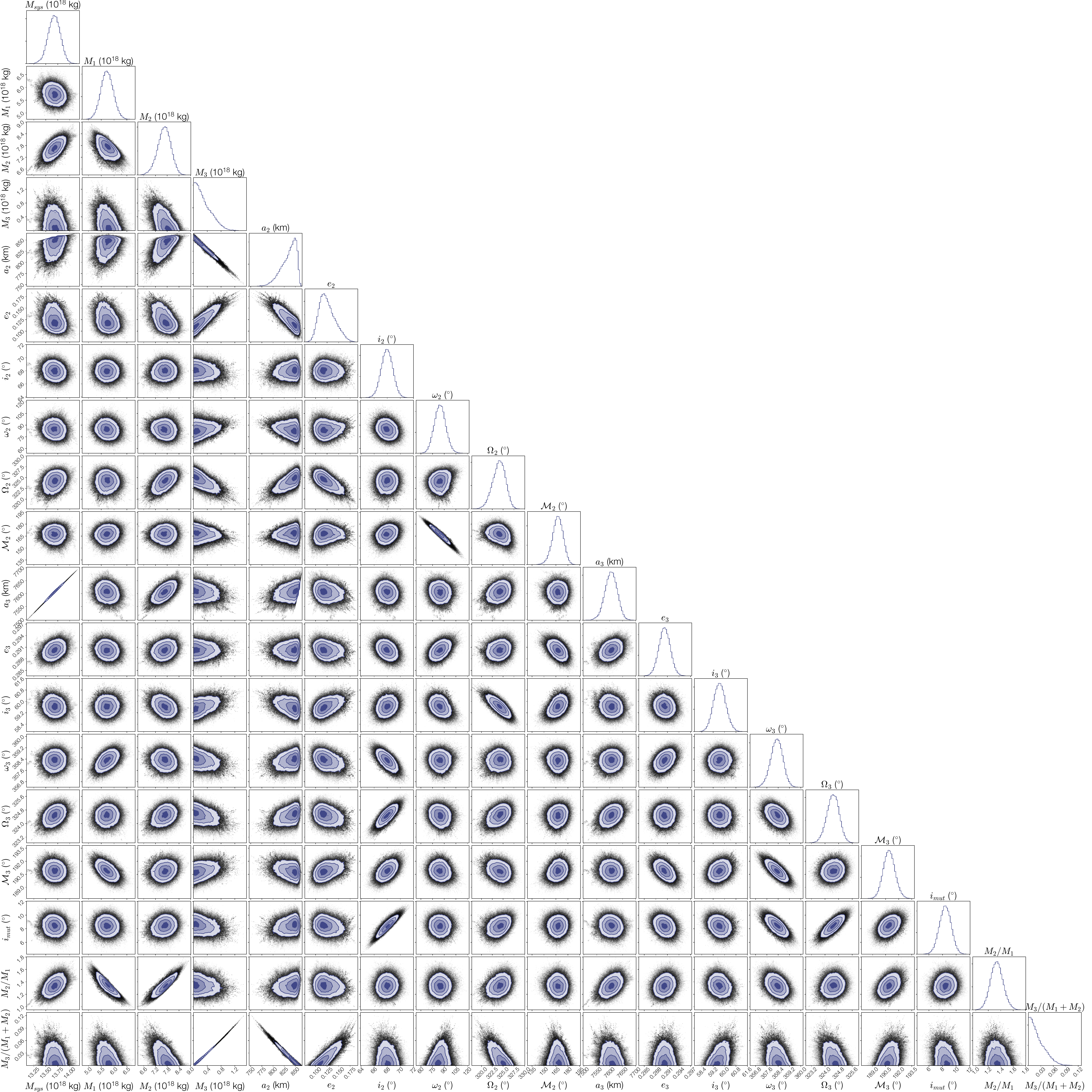}
    }
    \caption{A corner plot showing the joint posterior distribution of our 3-body fit using \multimoon{}. Histograms at the top of each column show the marginal distribution for each parameter. The joint posterior for each pair of parameters is shown in the contour plots, where the black points show sampled locations. Contours show the 1, 2, and 3$\sigma$ confidence intervals for each joint posterior distribution. All angles are relative to the J2000 ecliptic plane and all parameters are based on an epoch of JD 2453880.0 (2006 May 24 12:00 UT). The strongest degeneracies are for combinations of masses and semi-major axes that lead to equivalent orbital periods (such as $a_3$ and $M_{sys}$) and for combinations of angles that project to the same orientation in physical space (such as $\omega$ and $\mathcal{M}$). The mass of outermost component Paha ($m_3$) is consistent with zero and has a 95\% confidence upper limit of $\sim$0.03 of the total system mass. See Table \ref{tab:fit_results} for specific values. 
    }
    \label{fig:corner}
\end{figure}

Over the course of testing and producing the fit discussed in this paper, over $10^9$ model evaluations have been completed and compared to the data. This section discusses the results of our final fitting. These results are summarized in Table \ref{tab:fit_results}, showing the best fit, median, and confidence intervals of each fitted parameter with comparisons to \citet{benecchi201047171}.  We also display the joint posterior distributions of parameters in Figure \ref{fig:corner} using a corner plot \citep[][]{corner}. Throughout, we refer to 16$^{th}$ and 84$^{th}$ confidence intervals as ``1-$\sigma$'' for convenience of comparison, even though the posterior parameter distribution may not be Gaussian.

We found that our three-body fits give a reduced $\chi^2 = 2.38$ ($\chi^2 = 49.9$ with 21 degrees of freedom). The fitting in B10 gave $\chi^2 = 42.6$ for the inner binary alone, with a far worse $\chi^2$ value for the orbit of Paha. Clearly, the three-body model presented here is far superior to that of B10, but still has many deficiencies, as shown by the still high reduced $\chi^2$ value. We discuss the reasons for this below. 

We note the value of Bayesian parameter inference for the non-Keplerian solution as it is able to recover statistical upper limits on the mass of Paha ($M_3$), non-Gaussian distributions of parameters like the Lempo-Hiisi semi-major axis ($a_2$), and statistically rigorous distributions of derived parameters like the true three-dimensional mutual inclination between the inner and outer orbits ($i_{mut}$). Access to the posterior distribution is available upon request.

\begin{table}[]
    \centering
    \label{tab:fig_results}
    \begin{tabular}{lcccc}
    \hline
Parameter & & Posterior Distribution & Best fit & B10 fit \\ \hline
\textit{Fitted Elements} & & & \\
\qquad Mass Lempo ($10^{18}$ kg) & $M_1$ & $5.725^{+0.245}_{-0.235}$ & 5.960 & \nodata \\
\qquad Mass Hiisi ($10^{18}$ kg) & $M_2$ & $7.657^{+0.325}_{-0.347}$ & 7.610 & \nodata \\
\qquad Mass Paha ($10^{18}$ kg) & $M_3$ & $0.248^{+0.303}_{-0.176}$ & 0.070 & \nodata \\
\qquad Semi-major axis Hiisi (km) & $a_2$ & $838^{+13}_{-21}$ & 850 & $867\pm11$ \\
\qquad Eccentricity Hiisi & $e_2$ & $0.1230^{+0.0205}_{-0.0146}$ & 0.1084  & $0.101\pm0.006$ \\
\qquad Inclination Hiisi ($\degr$) & $i_2$ & $67.966^{+0.877}_{-0.877}$ & 68.229 & $68.8\pm0.6$ \\
\qquad Argument of periapse Hiisi ($\degr$) & $\omega_2$ & $84.6^{+7.3}_{-7.3}$ & 86.0 & $90.0\pm6.3$ \\
\qquad Longitude of the ascending node Hiisi ($\degr$) & $\Omega_2$ & $324.11^{+1.40}_{-1.58}$ & 324.89 & $327.6\pm6.3$ \\
\qquad Mean anomaly Hiisi ($\degr$) & $\mathcal{M}_{2}$ & $167.4^{+6.5}_{-6.8}$ & 165.7 & $136.7\pm1.6$ \\
\qquad Semi-major axis Paha (km) & $a_3$ & $7606^{+23}_{-24}$ & 7599 & $7411\pm12$ \\
\qquad Eccentricity Paha & $e_3$ & $0.2903^{+0.0015}_{-0.0015}$ & 0.2902 & $0.2949\pm0.0009$ \\
\qquad Inclination Paha ($\degr$) & $i_3$ & $59.652^{+0.431}_{-0.418}$ & 59.524 & $60.6\pm0.2$ \\
\qquad Argument of periapse Paha ($\degr$) & $\omega_3$ & $358.30^{+0.42}_{-0.44}$ & 358.53 & $342.0\pm0.2$ \\
\qquad Longitude of the ascending node Paha ($\degr$) & $\Omega_3$ & $324.47^{+0.35}_{-0.37}$ & 324.65 & $319.9\pm0.1$ \\
\qquad Mean anomaly Paha ($\degr$) & $\mathcal{M}_{3}$ & $190.90^{+0.68}_{-0.66}$ & 190.43 & $349.0\pm0.2$ \\
\textit{Derived Parameters} & & & & \\
\qquad System Mass ($10^{18}$ kg) & $M_{sys}$ & $13.677^{+0.127}_{-0.129}$ & 13.640 & $14.20\pm0.05$, $12.75\pm0.06$ \\
\qquad Hiisi-Paha mutual inclination ($\degr$) & $i_{mut}$ & $8.42^{+0.85}_{-0.87}$ & 8.71 & $10.7\pm0.8$ \\
\qquad Density Lempo (kg m$^{-3}$) & $\rho_{1}$ & $542^{+104}_{-116}$ & 566 & \nodata \\
\qquad Density Hiisi (kg m$^{-3}$) & $\rho_{2}$ & $925^{+181}_{-193}$ & 919 & \nodata \\
\qquad Density Paha (kg m$^{-3}$) & $\rho_{3}$ & $206^{+254}_{-152}$ & 58 & \nodata \\
\hline
\end{tabular} 
    \caption{Our \multimoon{} fit results compared to \citet{benecchi201047171}. Angles reference the J2000 ecliptic frame. Densities use radii values from \cite{mommert2012tnos} discussed in the text. In this table we use Lempo, Hiisi, and Paha to reference components A1, A2, and B in \citet{benecchi201047171}. Our parameters are primaricentric, which is not an exact comparison to the \citet{benecchi201047171} parameters for the outer orbit (subscript 3). }
    \label{tab:fit_results}
\end{table}

\section{Discussion}
\subsection{Model Fits}

The fits presented here represent the first measurement of the masses of each component in the Lempo system. \citet{benecchi201047171} estimated a \textit{system mass} of 12.75$\times 10^{18}$ kg or 14.20$\times 10^{18}$ kg for the system, based on the orbit fits of Paha and the inner Lempo-Hiisi binary respectively. Our fitting gives a system mass of 13.630$\times 10^{18}$ kg, intermediate between the two values given in B10. We also find that Hiisi (component A2 in \citet{benecchi201047171}) is more massive than Lempo (component A1), which is somewhat surprising since it is $\sim$10\% fainter on average.\footnote{Throughout this paper, we have used the names Lempo and Hiisi to correspond to components A1 and A2 in B10. This is in correspondence with the determination of \citet{mommert2012tnos} of the diameters of the objects and the Minor planets center circular naming the objects. As discussed earlier, these measurements are based on assumed equal albeoes that may be invalid for Lempo and Hiisi. Though the MPC Circular states that the largest object is named Lempo with the second largest being Hiisi, we propose that the typically brighter component A1 be referred to as Lempo and the typically fainter component A2 be referred to as Hiisi, as we have done in this paper.}.  While our fits do not detect the mass of Paha, we estimate an upper limit of $\sim$0.5$\times 10^{18}$ kg. Given the radiometric sizes of Lempo, Hiisi, and Paha reported in \citet{mommert2012tnos}, we find the densities of each object (assuming spherical shapes and equal albedos) to be $\rho_1$ = 0.544 g cm$^{-3}$, $\rho_2$ = 0.929 g cm$^{-3}$ and $\rho_3 <$ 0.460 g cm$^{-3}$ (1-$\sigma$). The $\sim$2.5-$\sigma$ density difference between Lempo and Hiisi is somewhat concerning, but given the assumptions underlying the estimates, is not too worrying. Small differences in albedo, shapes, lightcurves, or thermal properties could affect these measurements significantly. 
Despite these caveats, it is clear that all three bodies in this system are extremely porous, in line with density measurements of other TNBs at this size. 

The low densities along with the nearly coplanar orbits of the three components are indicative of formation by the streaming instability. \citet{nesvorny2010formation} suggested that coplanar triple systems are a somewhat common outcome of simulations of formation of TNOs via graviational collapse initiated by the streaming instability. Alternatively, binary-binary interactions during which a collision or capture takes place would in theory be able to recreate Lempo-like systems \citep[][]{brunini2020origin}, but these produce systems with a more random inclination distribution. With precise characterization of Lempo (and other trinary systems yet to be discovered), there is an opportunity to calibrate these planet formation models. 

\citet{correia2018chaotic} showed that, based on the orbit parameters found in B10, the spin-orbit dynamics in the Lempo system are highly chaotic. While we found slightly different parameters than B10, we expect that the analysis of \citet{correia2018chaotic} is still valid since we continue to find a surprisingly high eccentricity of $0.12^{+0.02}_{-0.01}$ (and the same orientation) of Lempo-Hiisi. Unlike the expectation of \citet{correia2018chaotic} that a more precise analysis would reveal a low eccentricity, our work confirms that the highly chaotic solution of \citet{benecchi201047171} is still valid! Note that \citet{correia2018chaotic} established not only that the Lempo system was chaotic, but that this chaos led to instability and system disintegration within 100 kyr even when including the effect of tidal damping. 

The mystery of the high eccentricty of Lempo-Hiisi can't easily be explained. For example, perhaps the three-point-mass model is inadequate. Spin dynamics are expected to be very strong in this system, so perhaps a model that includes spin dynamics would result in a more accurate and smaller eccentricity measurement that is non-chaotic. We discuss below preliminary analyses that include the $J_2$ of Lempo and Hiisi, but these still suggest a high eccentricity that is not significantly different than \citet{benecchi201047171}. Additional analyses with additional data are warranted. The inner Lempo-Hiisi binary can be resolved from the best ground-based telescopes under excellent observing conditions or potentially by JWST NIRCam in the shortest wavelength (with similar resolving power to the now defunct HST High Resolution Camera that was essential for precise astrometry of the tight inner binary). Perhaps the relative astrometry for these highly-blended observations is not fully accurate or well-explained by the reported Gaussian uncertainties. 

Another possibility is that the Lempo components all spin rapidly, since \citet{correia2018chaotic} found that spins of $\lesssim$ 6 hours tend to avoid spin-orbit chaos. This seems inconsistent with lightcurve observations as summarized by \citet{benecchi201047171}. Additionally, \citet{correia2018chaotic} used a mass for Paha higher than the upper limit that our fit places. It is possible that instability is avoided if Paha has a sufficiently small mass.

While solving the mystery of Lempo is an important scientific goal in and of itself, we also note that the Lempo system is a touchstone for understanding the formation of the outer solar system overall. First, Lempo has a unique architecture as a ``trinary system'' of three similarly-sized bodies. Second, whatever the resolution of the dynamical mystery of its ``stable chaos'', the complexity of the system means that more can be learned about Lempo than in other systems; for example, it may be a rare system where tidal dissipation parameters can be meaningfully constrained. Third, the formation of such trinaries is generally predicted by streaming instability models \cite[e.g.,][]{nesvorny2021binary}, but using the configuration of these triples to constrain the models of gravitational collapse has hardly been explored. Finally, Lempo has a long history of observational data, going back to 2001. No other trinary will ever have as long of an observing baseline as Lempo! The spin-orbit dynamical effects of interest operate on decadal timescales making Lempo a unique window into these effects.

The reduced $\chi^2$ value of the best fit orbit show that the three-body fitting completed here is still inadequate in fully describing the dynamics of the Lempo-Hiisi-Paha system (assuming the uncertainties on the astrometric positions are accurate). As the inner binary is extremely close ($q/R$ $\sim$5.5), we expect that non-Keplerian effects from the shapes of Lempo and Hiisi will be very strong. Using a reasonable estimate of $J_2 = 0.05$, we estimate that both the nodal and apsidal precession from non-Keplerian effects could be $\sim$0.5$\degr$ day$^{-1}$ for the inner binary \citep[][]{scheeres2000evaluation}. Assuming a three-point-mass fit returns a nodal angle with an uncertainty of only 1.5$\degr$ despite the data spanning $\sim$1500 days. Clearly, modeling this spin-induced precession is crucial in fully understanding the system. 

In addition to three-body orbital fits, we also attempted several fits where the $J_2$ of Lempo and Hiisi was added. The best fit from these runs had a similar fit quality to our best fit three-body solution. Although we are achieving similar fit quality, full convergence of the MCMC chains would require at least an order of magnitude more computational effort. As there are additional ground-based observations of Lempo available, we defer to a future analysis a more detailed investigation. 

Even without converged MCMC chains, however, the best fits from these preliminary runs show that inclusion of non-Keplerian shape effects are justified. Our best fits from these runs have somewhat large values for the $J_2$ of Lempo and Hiisi, with $J_2 \sim 0.1$ for both objects. Additionally, the spin poles of Lempo and Hiisi are fairly well aligned with their mutual orbit. The fits also show that most orbital parameters of the binary are similar to the values found in the three-body fit presented above, including the eccentricity of the Lempo-Hiisi binary and the masses of Lempo and Hiisi. The mass of Paha, however, seems to be higher than would be expected based on the three-body fit. This is not unsurprising, however, since the mass of Paha and $J_2$ both cause precession of the inner binary. These insights, although very preliminary, show that much could be learned from more complex modeling of the Lempo system. 



\section{Conclusions}
\label{sec:conclusions}
Unlocking the full potential of SSSBs to inform theories of the formation and evolution of the solar system requires moving beyond point masses and studying non-Keplerian effects. After careful consideration, we discuss how the next generation of dynamical analyses should include non-Keplerian effects from the quadrupolar gravitational potentials (characterized by \jt{}, \ct{}, and spin orientation), the Sun, and unknown companions. We develop and share with the community a new $n$-quadrupole integrator, \spinny{}, which can include these effects and is optimized for short integrations relevant to observational timescales. \spinny{} is the integrator at the core of a new Bayesian parameter inference modeling tool, \multimoon{}, which efficiently produces posterior probability distributions for orbital, shape, and spin properties based on relative astrometry data common for SSSBs. By design, \multimoon{} is user-friendly and open-source. We highlight the assumptions in these new tools and their validation. 

We then apply \spinny{} and \multimoon{} to the long-standing challenge of a more advanced model of Lempo trinary system. We find the best three-point-mass model of this system and provide probability distributions for all of its parameters in Table \ref{tab:fit_results} and Figure \ref{fig:corner}. For example, the fainter Hiisi is about 33 $\pm$ 5\% more massive than Lempo. Surprisingly, our solution was similar to the \citet{benecchi201047171} results that  \citet{correia2018chaotic} showed to be highly chaotic and unstable on timescales $\sim$10$^5$ times shorter than the age of the solar system. Possible resolutions to this conclusion (including quadrupoles or rapid spins) seem inconsistent with the data, but further analysis is required. 

Non-Keplerian analyses will soon be essential to precise modeling of SSSBs in general as these effects become significant compared to the observational uncertainties on decadal timescales (Proudfoot et al., in prep.). Use of tools like \spinny{} and \multimoon{} will enable the community to gain new scientific insights about the shapes and spins and configurations of SSSBs as our models go beyond point masses.

\begin{acknowledgments}
This research is based on observations made with the NASA/ESA Hubble Space Telescope obtained from the Space Telescope Science Institute, which is operated by the Association of Universities for Research in Astronomy, Inc., under NASA contract NAS 5–26555. Development of \spinny{} and \multimoon{} were supported by Programs HST-AR-14581 and HST-GO-15460. This work was also supported by grant 80NSSC19K0028
from NASA Solar System Workings. We thank William Giforos, Joseph Henderson, Maia Nelsen, and Meagan Thatcher for support in validating and testing \multimoon{}. 
\end{acknowledgments}

\bibliographystyle{apj}
\bibliography{all}

\end{document}